\documentclass[prl,twocolumn,letterpaper,subscriptaddress,amsmath,amssymb]{revtex4}
\usepackage[pdftex]{graphicx}
\graphicspath{{figures/pdf/}}
\DeclareGraphicsExtensions{.pdf}
\usepackage{dcolumn}% Align table columns on decimal point
\usepackage{bm}% bold math
\usepackage{amsthm}

\def\op#1{\hat{#1}}
\def\vec#1{{\bf#1}}
\def\op#1{\hat{#1}}
\def\op#1{#1}
\def\ket#1{| #1 \rangle}
\def\bra#1{\langle #1 |}

\def\ave#1{\langle #1 \rangle}

\def\dim{\operatorname{dim}}

\def\F{\mathcal{F}}

\def\eps{\epsilon}
\def\opt{\rm opt}
\def\Im{\operatorname{Im}}

\begin{document}
\title{Fast, high fidelity information transmission through spin chain
quantum wires}
\author{S.~G.~Schirmer}\email{sgs29@cam.ac.uk}
\affiliation{Department of Applied Maths and Theoretical Physics,
             University of Cambridge, Wilberforce Road, Cambridge, CB3 0WA, UK}
\author{P.~J.~Pemberton-Ross}\email{pjp32@cam.ac.uk}
\affiliation{Department of Applied Maths and Theoretical Physics,
             University of Cambridge, Wilberforce Road, Cambridge, CB3 0WA, UK}
\date{\today}

\begin{abstract}
Spin chains have been proposed as quantum wires for information transfer
in solid state quantum architectures.  We show that huge gains in both
transfer speed and fidelity are possible using a minimalist control
approach that relies only a single, local, on-off switch actuator.
Effective switching time sequences can be determined using optimization 
techniques for both ideal and disordered chains.  Simulations suggest
that effective optimization is possible even in the absence of accurate
models.
\end{abstract}
\maketitle

The dynamical evolution of a chain of coupled spins is well-suited to
the job of a data bus, transferring quantum information over the short
distances between quantum registers.  This is particularly desirable
in architectures where internal quantum communication is required but
the use of photons is impractical~\cite{NAT417p709,PRL90n087901}.  A
practical scalable data-bus should achieve fast, high fidelity of
information transfer without the need for intricate manipulation of
its parts.  Ideally, the data should flow along the bus without any
external control~\cite{BoseRev}.  However, without any dynamic control
perfect information transfer for most spin chains is possible only if
the couplings are precisely
engineered~\cite{PRL93n230502,PRA72n030301,PRA73n032306}.  Although
it often suffices to engineer the couplings near the end of the
chain~\cite{PRA72n034303}, fabrication of chains with such precisely
engineered couplings is a significant challenge.  Perfect state
transfer for any length of chain is also possible in principle if we
have full dynamic control over individual couplings between adjacent
spins, using either dynamic or adiabatic passage
schemes~\cite{qph0702019,NJP9p155}.  However, these schemes require
control of all couplings, and thus multiple control electrodes, which
aside from increasing system complexity may also be sources of noise
contributing to decoherence, an effect potentially amplified by the
high field strengths and longer transfer times required for adiabatic
passage.

Recent work has shown that many systems are controllable even if we
can only modify the dynamics of a small part of the system.  In
particular, the dynamics on the first excitation subspace of many spin
chains is controllable if we can vary the coupling at one end of the
chain~\cite{PRA78n062339}, for instance.  It has also been shown that
this global controllability can be exploited to improve the transfer
fidelities in certain types of spin chains by repeatedly applying
certain unitary operations at one end of the
chain~\cite{PRA75n062327}, and in the limit of instantaneous gate
operations and fast repetition, it can be shown that arbitrarily high
transfer fidelities can be achieved.  Theoretical controllability
results, however, suggest that one could do much better.  Almost any
local perturbation of the Hamiltonian theoretically suffices to
effectively control the system, and any objective can be achieved in
finite time by simply switching this perturbation on and off at
specific times.  In this Letter we present a systematic approach to
finding the correct switching times to achieve fast high fidelity
information transfer for various spin chains, including both uniform
and disordered spin chains.  We further demonstrate that effective
bang-bang switching sequences can be found even in the absence of a
precise model of the system, using adaptive closed-loop experiments.
This is important as the Hamiltonians for a particular physical
realization of a spin chain quantum wire are usually at best
approximately known and subject to variation due to fabrication
tolerances, although this problem can potentially be overcome by
experimental system identification~\cite{PRA79n020305}.

%\cite{PRB71n100501, NJP7p181,NJP9p79,PRL91n207901,JOB7pS356,qph0606188,CM16p4991,MJ37p1440}

%%% system description
For proof-of-principle simulations we consider spin chains of $N$
spin-$\frac{1}{2}$ particles with a coupling Hamiltonian 
\begin{equation}
    \op{H}_I = \frac{1}{2} \sum_{n,m} J_{mn}^x \sigma_m^x \sigma_n^x
                        + J_{mn}^y \sigma_m^y \sigma_n^y
                        + J_{mn}^z \sigma_m^z \sigma_n^z,
\end{equation}
where $\sigma_n^\ast$ for $\ast\in\{x,y,z\}$ are the usual Pauli
matrices for the $n$th spin and $J_{mn}=J_{nm}$ are the coupling
constants.  This model covers common types of chains, including XY,
Heisenberg, and dipole-coupled spin chains.  For applications such as
information transfer, it is desirable to restrict the dynamics by
ensuring that $H_I$ commutes with the total spin operator
$S=\sum_{n=1}^N\sigma_n^z$, and thus conservation of the total number of
excitations.  This condition is equivalent to $xy$-isotropy, i.e.,
$J_{mn}^x=J_{mn}^y$, and ensures that the Hamiltonian decomposes into
excitation subspaces.  As usual in work on spin chains for information
transfer we assume $J_{mn}^x=J_{mn}^y$ and work in the first excitation
subspace $E_1$.  We have $\dim E_1=N$ and the off-diagonal elements of
the Hamiltonian restricted to $E_1$ are $H^{(1)}_{mn}=J_{mn}^x=J_{mn}^y$
for $m\neq n$.  The diagonal elements depend on $J_{mn}^z$.  Most
simulations will be based on isotropic XY or Heisenberg chains, where
$J_{mn}^x=J_{mn}^y$ and $J_{mn}^z=0$ in the former case, and
$J_{mn}^x=J_{mn}^y=J_{mn}^z$ in the latter.  For chains with
nearest-neighbor coupling (nnc) $J_{mn}=0$ except for $n=m\pm 1$.  An
nnc chain is uniformly coupled (unnc) if $J_{n,n+1}=J$ for all $n$, and
choosing time in units of $J^{-1}$, we may assume $J=1$.

\begin{figure}
\includegraphics[width=0.48\textwidth]{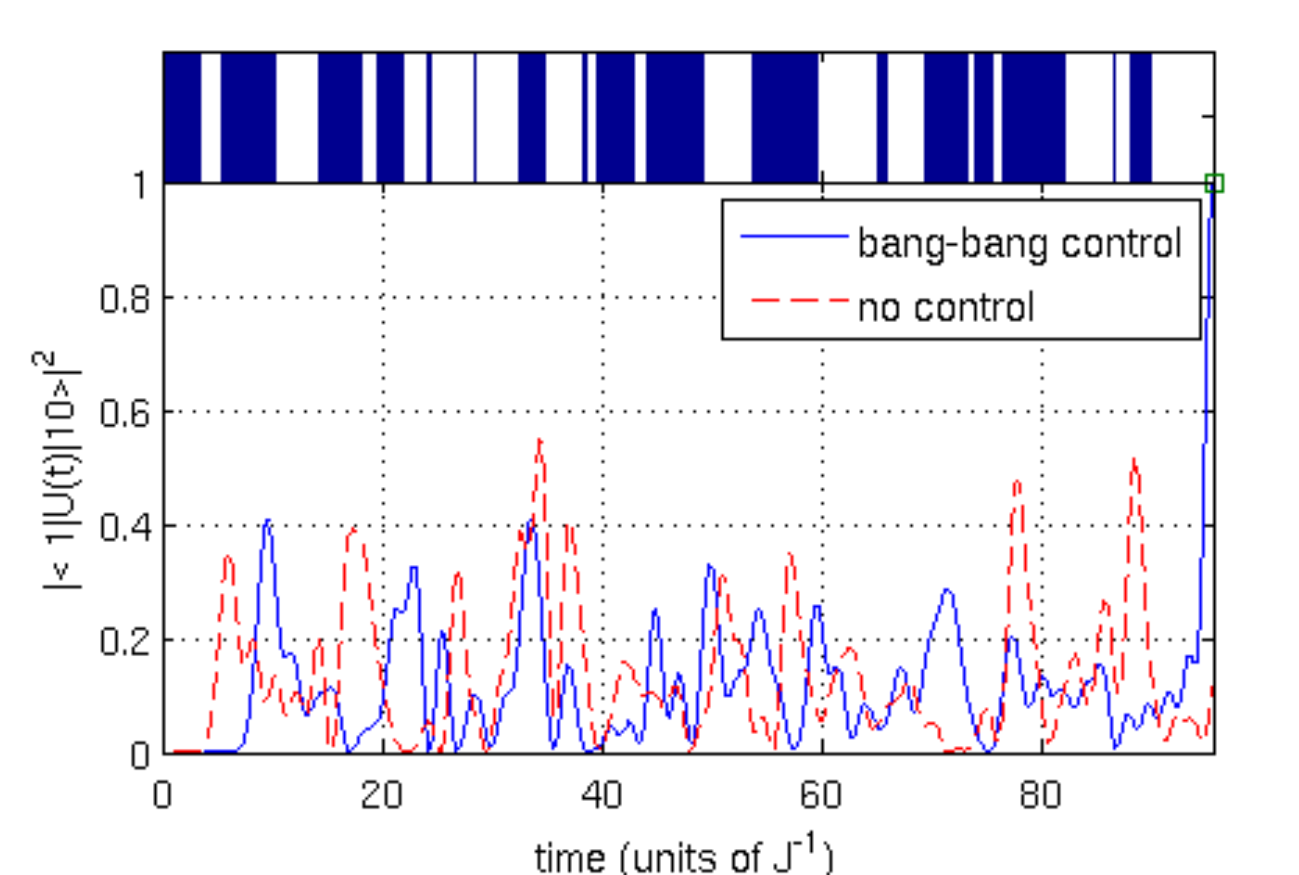}
\caption{Population of $10$th spin for length-$10$ disordered nnc XYZ
chain without control (red) and with optimized bang-bang control (blue).
Although the actuator only effects a local perturbation of the coupling
between spins 1 and 2, it significantly changes the evolution of the
entire system, resulting in ``constructive interference'' of the
 populations at the $N$th spin at the target time $T=95.4740$.}
\label{fig:flow}
\end{figure}

%%% control and optimization objectives
The objective of optimizing information transmission through a spin
chain quantum wire is to achieve high-fidelity transfer in a minimum
amount of time.  We aim to achieve this by optimizing the switching 
time sequence $\vec{t}=(t_1,\ldots,t_K)$ of a simple binary switch 
actuator, such as a local control electrode.  The evolution of the 
system subject to this bang-bang control is given by
\begin{equation}
  \label{eq:Uop}
  U(\vec{t}) = U^{(1)}(t_1) U^{(2)}(t_2) \ldots U^{(1)}(t_{K-1}) U^{(2)}(t_K),
\end{equation}
where $U^{(m)}(t_k)=\exp(-i t_k \op{H}_m)$.  $U^{(1)}(t_k)$ corresponds
to free evolution under $H_1=H_I$ (actuator off), and $U^{(2)}(t_k)$ to
evolution under the perturbed Hamiltonian $H_2=H_I+H_C$ for $t_k$ time
units.  The idea is that although the actuator changes the Hamiltonian
only locally in a fixed way, by switching this perturbation on and off
at different times, we can realize basically infinitely many effective
Hamiltonians that give rise to different evolutions of the system.  Most
of these will not produce a desirable evolution, but some are likely
have desirable characteristics such as a near unit-height peak in the
population of a particular state after a short time as shown in
Fig.~\ref{fig:flow}.  The optimization tries to find such desirable
cases by systematically exploring the parameter space, in our case the
possible switching time sequences $\vec{t}$.  Specifically, we seek
$\vec{t}$ that minimize the average transmission error for a quantum bit
propagating through the chain~\cite{BoseRev}, $E(\vec{t}) =
1-|\bra{N}U(\vec{t})\ket{1}|^2$, subject to the constraint that the
transmission time $T=\sum_{k=1}^K t_k$ and number of switches remain
below thresholds $T_{\rm max}$ and $K_{\rm max}$.

%%% basic optimization routine

If the chain Hamiltonian $H_I$ and the perturbation $H_C$ induced by the
actuator are known then we can easily solve the optimization problem.
For fixed $K$ the gradient of the objective function is $\nabla E =
(\partial_1 E,\ldots,\partial_K E)$, where the partial derivatives are
\begin{equation} \textstyle
 \partial_k E = \frac{\partial E}{\partial t_k}
  = -2\Im \left[\bra{N} U^{(k)}(\vec{t})\ket{1}
                \bra{1}U(\vec{t})^\dag\ket{N}\right]
\end{equation}
with $U^{(k)}(\vec{t})= U_1 \cdots H_k U_k \cdots U_K$, where $U_\ell$
is the $\ell$th factor in~(\ref{eq:Uop}).  Equipped with this gradient
information we can use either a gradient descent algorithm to find 
$\vec{t}_{\rm opt}$, or we can similarly calculate the Hessian matrix 
of 2nd derivatives $H=(H_{k\ell})$ with
$H_{k\ell}=\partial_k\partial_\ell E(\vec{t})$,
and use the more efficient Newton method with an adjustable step size 
$\gamma>0$ to iteratively update $\vec{t}$
\begin{equation}
  \vec{t}_{s+1} = \vec{t}_s - \gamma [H(E(\vec{t}_s))]^{-1} \nabla E(\vec{t}_s)
\end{equation}
until convergence is achieved.  In practice there are some additional
complications as we have constraints on the switching times (e.g.
$t_k\ge t_{\min}$, $\sum_k t_k \le T_{\max}$, etc) but these can easily
be incorporated in the algorithm~\cite{JOTA22p297}.  The resulting
$\vec{t}_{\opt}$ depends on the initial guess $\vec{t}_0$, and as with
virtually all optimization algorithms, we can only guarantee convergence 
to a local optimum, but the method is generally both effective and very
efficient.

%%% vary chain length 

\begin{figure}
\includegraphics[width=0.48\textwidth]{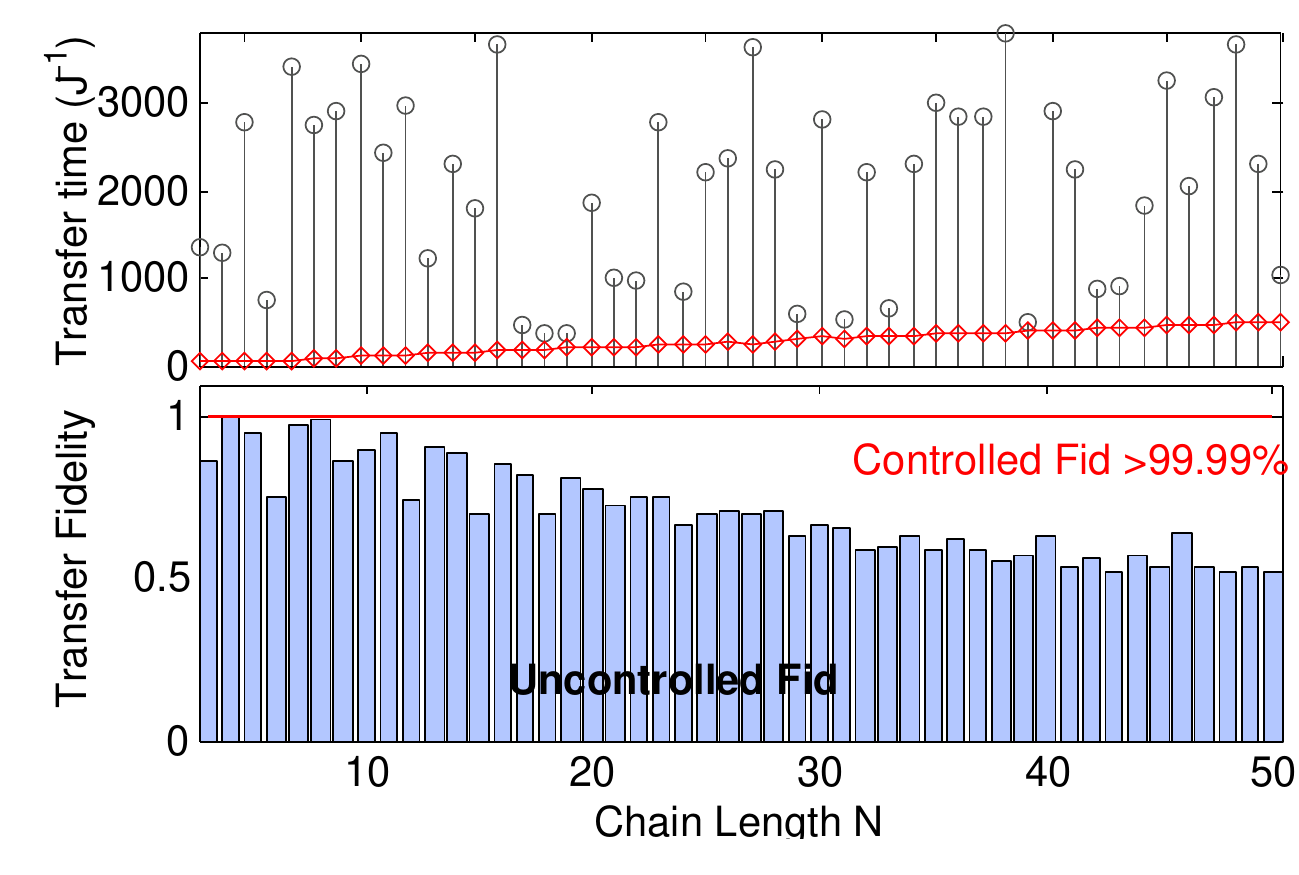}
\caption{Transfer times and fidelities for Heisenberg spin chain 
with uniform nearest neighbor coupling.  Best fidelities achievable
according to in at most 4000 time units without control (blue bars /
circles) and with optimized bang-bang control (red diamonds/line).}
\label{fig:Bose}
\end{figure}

We first tested if we are indeed able to achieve fast, high-fidelity
information transfer using a simple binary switch actuator for unnc XY
and XYZ chains.  For such chains perfect information transfer without
control or engineered couplings is possible only for $N=3$, and as
Fig.~\ref{fig:Bose} for the Heisenberg chain shows, the maximum transfer
fidelity in a limited amount of time decreases quickly, and the peak
transfer times vary erratically with the chain length $N$.  Assuming a
simple local actuator that switches off the coupling between the first
two spins, we were able to find switching sequences $\vec{t}$ that
achieved transfer fidelities $>0.9999$ for both XY and XYZ chains up to
length $50$ with transfer times $T(N)\approx 10 N$ and at most $K=4N$
switches.  Even when the precision of the switching times $t_k$ was
limited to $10^{-4}$, i.e., four decimal digits, the error remained
below threshold except for XYZ chains with $N=48$, $50$, where the error
was slightly above threshold ($1.7\times 10^{-4}$ and $5.6 \times
10^{-4}$).  Surprisingly, we were able to achieve transfer fidelities
$>0.9999$ in many cases even when the switching times were limited to
only three decimal digits accuracy, and for shorter chains even fewer
digits appeared sufficient in many cases, though at the expense of
somewhat increased transfer times.  This is significant as it shows that
we do not require ``double precision'' control of the switching times,
which would be experimentally infeasible.

%%% KT optimization

\begin{figure}
\includegraphics[width=0.48\textwidth]{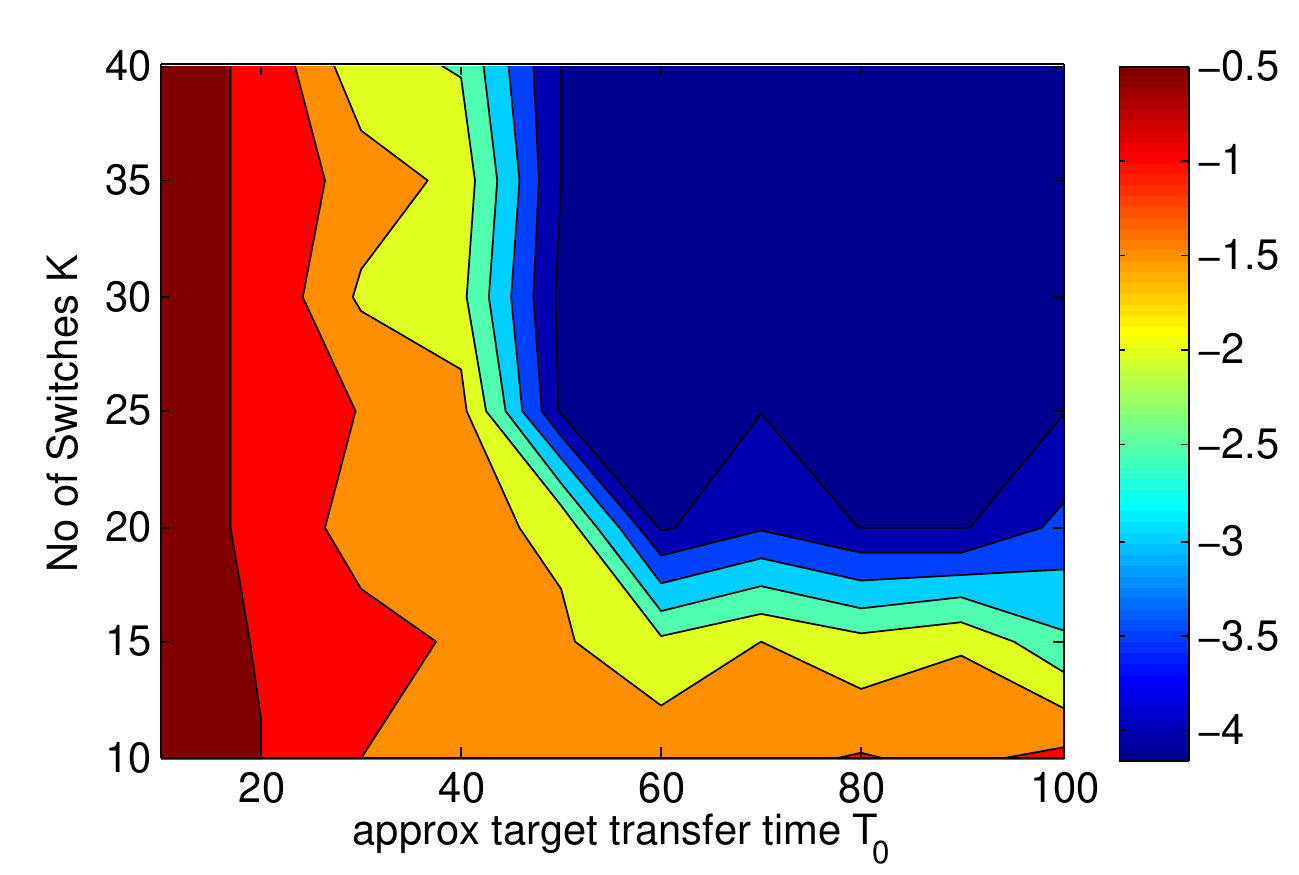}
\caption{Error $\log_{10}(E(\vec{t}_{\opt}))$ as a function of the
number of switches and target transfer times $T_0$, where
$E(\vec{t}_{\opt})$ is the minimum error over 10 runs with random
initial $\vec{t}_0$} \label{fig:KT}
\end{figure}

Although the optimal transfer times of $T\approx 10N$ are impressive
compared with those for the uncontrolled chain, especially considering
the transfer fidelities, the ease with which the algorithm appeared to
be able to find solutions suggested that both $K$ and $T$ could be
further reduced.  To investigate this, we systematically varied the
number of switches and approximate target transfer times $T_0$ for a
benchmark case of an XYZ unnc chain of length $10$.  The contour plot of
the error on a logarithmic scale ($\log_{10} E(\vec{t}_{\rm opt})$) as a
function of $K$ and $T_0$ (Fig.~\ref{fig:KT}) suggests that for this
system near-perfect transfer could be accomplished in as little as 50
time units (measured in units of $J^{-1}$) with as few as $22-24$
switches.  Pushing the limits tends to slow down the optimization,
increasing the number of iterations and possibly requiring several runs
with different initial guesses $\vec{t}_0$.  Yet, for the simple
benchmark problem, all runs completed in a few seconds (average $1.3$s,
max.~4.6s) on a standard laptop.

%%% perturbed chains

\begin{figure}
\includegraphics[width=0.48\textwidth]{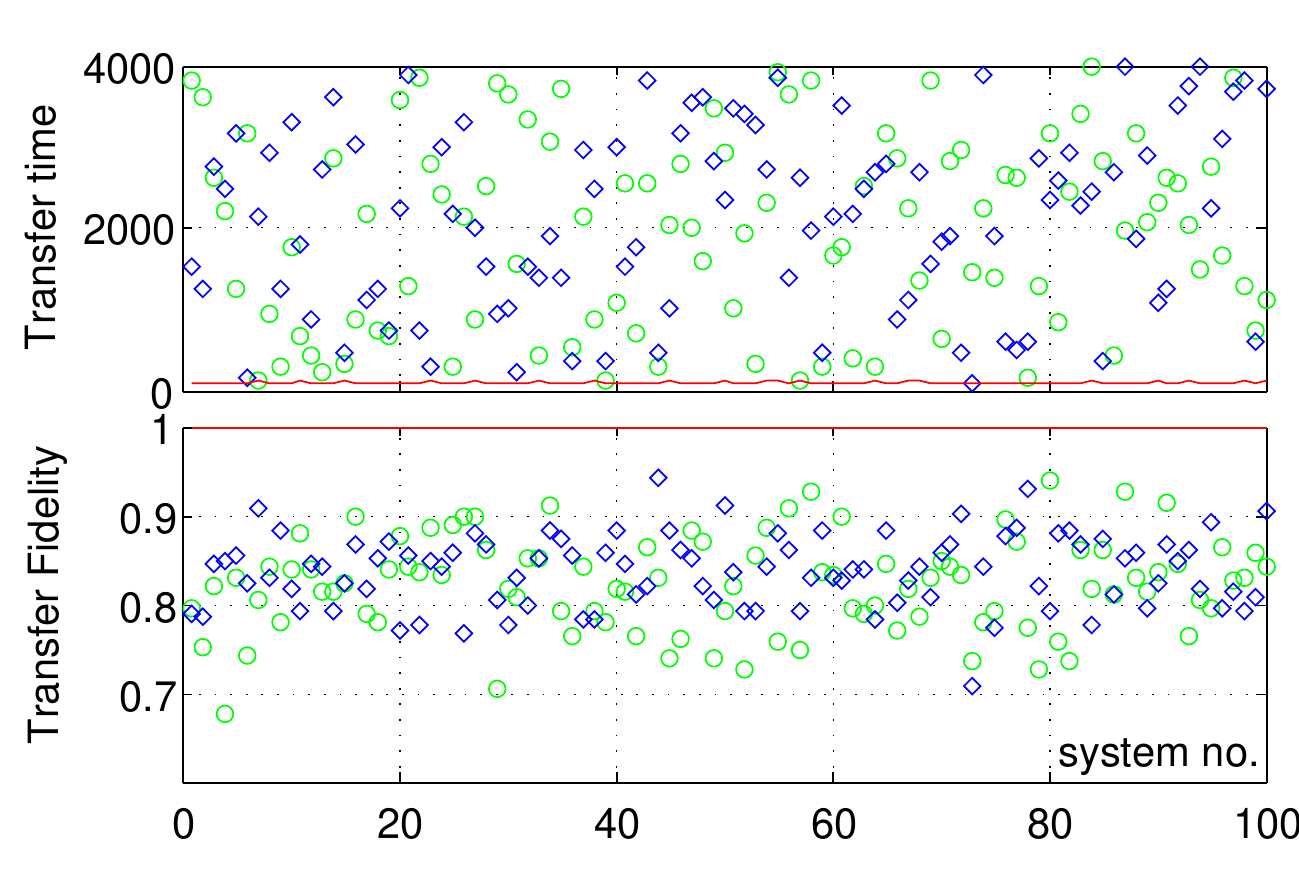}
\caption{Transfer times $T$ and fidelities $\F$ for 100 (nnc) Heisenberg
chains of length $10$ with randomly perturbed couplings
$J_{n,n+1}=1+\eps \xi$ ($\xi$ Gaussian random variable, $\ave{\xi}=0$,
$\sigma(\xi)=1$) for 1\% and 10\% disorder.  Without control, the
maximum transfer fidelities achievable in $\le 4000$ time units and
transfer times $T$ are scattered with $\ave{\F}=83.74$\%,
$\ave{T}=2,120$ for $\eps=0.01$ (blue diamonds) and $\ave{\F}=82.21$\%,
$\ave{T}=1,956$ for $\eps=0.1$ (green circles).  With optimized
bang-bang control, near-perfect transfer $\ave{\F}\ge 0.9999$ for
$\eps=0.01$ and $\ave{\F}\ge 0.9995$ for $\eps=0.1$ can be achieved for
all 100 test systems in approximately 100 time units (red lines).}
\label{fig:Perturbed}
\end{figure}

While chains with uniform nearest neighbor coupling provide nice models,
in practice the coupling constants $J_{mn}$ are likely to be subject to
variation, and direct coupling may not be limited to nearest neighbors.
It is thus crucial to investigate the performance of the bang-bang
control scheme and optimization algorithm for perturbed chains.  As
Fig.~\ref{fig:Perturbed} shows, with very few exceptions the algorithm
had no difficulty in finding solutions that achieved threshold transfer
fidelities of $99.99$\% with transfer times of $T\approx 100$ and $K\le
40$ switches for moderately disordered nnc Heisenberg chains of length
$10$.  Similar results were achieved for other chains, e.g. XY chains.
Preliminary simulations suggest that substantial gains in fidelities and
transfer times are possible even for highly disordered chains, although
in some cases finding solutions becomes challenging and target transfer
times and/or the number of switches may need to be increased.

%%% closed-loop optimization

These results are promising in that the simulations suggest substantial
improvements of both transfer fidelities and transfer times are possible
with very limited control and very simple actuators, even for disordered
chains.  However, our optimization procedures assumed knowledge of the
chain's Hamiltonian and the perturbation induced by the actuator. A
valid objection to the practical feasibility of the approach is that
this information is often simply not available for a particular physical
system.  Furthermore, unknown environmental effects may perturb the
evolution.  Hence, it is crucial to consider if we can determine a set
of optimal switching times for a given system without recourse to an
idealized model, using adaptive closed-loop experiments.  In this case,
the figure of merit, in our case the transfer fidelity, for a particular
switching sequence is evaluated experimentally as follows: initialize
the system, create an excitation at one end, apply the control sequence,
the measure of spin at the other end of the chain.  Assuming a simple
binary-outcome projective measurement, this experiment is repeated until
we have accumulated sufficient data to estimate the transfer fidelity to
a desired number of significant digits.  In this setting the
optimization must find an optimal time sequence based only on the
limited-precision fidelity measurement data.

To assess if we can still find effective switching pulse sequences we
simulated this situation, i.e., we choose a model system to calculate
the fidelities, but instead of providing the optimization routine with
information about the actual model as before, it now only had access to
an estimate for the fidelity with $D$ significant digits for each time
sequence $\vec{t}$.  For our benchmark problem (a unnc XYZ chain of
length $10$) we compared for four types of algorithms: genetic, pattern
search, Nelder-Mead simplex and Newton iteration with discrete gradients
derived from the limited-precision (simulated) measurement data.  The
results are shown in Table~\ref{table1}.  Pattern search was abandoned
due to extremely poor performance and convergence issues.  Most notably,
the standard genetic algorithm, despite being a popular choice for 
closed-loop optimization experiments in some areas, failed completely 
for this problem.  Out of 100 trials with different initial populations 
(50 individuals), not one came close to reaching threshold fidelity, 
and the number of fidelity evaluations (\#$\F$), and hence experiments 
required, was huge.  The simplex algorithm did significantly better in 
that 75\% of the trials succeeded in finding solutions above threshold 
fidelity.  Application of the quasi-Newton-type optimization procedure 
outlined above, requires that the analytical gradients be replaced by 
discrete gradients calculated from the fidelity measurements.  Although
the limited precision is a challenge here, with careful tuning of the
parameters in the discrete gradient estimation, the quasi-Newton routine
(Newton$^1$) far outperformed all other choices in terms of success
rate, number of fidelity evaluations, execution times, and best transfer
times.  For comparison we have included results for our model-based
Newton iteration (Newton$^2$) using (partially) analytic gradients.  As
expected, the model-based iteration is more efficient in terms of
function evaluations and execution times, but there is no difference in
the success rates or average transfer times.

\begin{table}
\begin{tabular}{|l||c|c|c|c|c|}
\hline
 & \% Success & $\ave{\# \F}$ & $\ave{t_{\rm exe}}$ & $\ave{T}$ & $T_{\min}$ \\\hline 
 genetic    &   0 & 12,275.5 & 35.4497 & 71.6552 & --- \\\hline 
 simplex    &  75 &  8,696.3 & 21.7677 & 99.9145 & 94.9778 \\\hline
 Newton$^1$ & 100 &  1,379.8 &  3.6295 & 99.5020 & 74.6144 \\\hline
 Newton$^2$ & 100 &     96.3 &  0.7087 & 99.6101 & 74.6268 \\\hline
\end{tabular}
\caption{Performance of various optimization algorithms in closed-loop
experiment simulations for benchmark problem.}
\label{table1}
\end{table}

%\begin{figure}
%\includegraphics[width=\columnwidth]{ClosedLoop}
%\caption{Adaptive closed-loop optimization: The system is initialized,
%the control sequence applied via the local actuator (red), and the
%transfer fidelity measured experimentally.  The measured fidelities are
%fed directly into an optimization algorithm and used to generate new
%switching sequences.  No Hamiltonian/model is required.}
%\label{fig:closed-loop-exp}
%\end{figure}

As estimating the fidelities to a large number of significant digits is
costly in that it requires many experiment repetitions for each fidelity
evaluation, we further investigated the effect of limiting the precision
of the fidelity measurements to a few significant (decimal) digits.  We
note that $4$ digits is the minimum accuracy required if we wish to
achieve transfer fidelities $\ge 99.99$\%.  We found that with suitable
choice of the algorithmic parameters, especially with regard to the
discrete gradient estimation, the quasi-Newton-type algorithm generally 
continued to find solutions with error probabilities $\le 10^{-4}$ in 
almost all cases even when the accuracy of the fidelities was limited to 
$5$ to $8$ digits, and in some cases the algorithm was able to find time 
sequences achieving transfer fidelities $\ge 0.9999$ for our benchmark 
problem even when both the accuracy of the fidelity measurements and the
switching times $t_k$ were limited four decimal digits.  This suggests
that threshold fidelities can be achieved even with limited precision
measurements and control.

%%% open problems and future work

We have shown that minimal control using a binary switch actuator that
induces a local perturbation to a fixed Hamiltonian holds considerable
promise to improve information transmission through spin chain quantum
wires.  The effectiveness of the technique for disordered chains and
the possibility of model-free optimization using closed-loop
experiments with limited precision measurements further enhance its
potential appeal.  Furthermore, the technique is not limited to
information transfer in spin chains with nearest neighbor coupling.
Preliminary simulations suggest that the technique is still effective
for more complex spin networks with non-nearest neighbor couplings and
potentially many other types of systems, and that it can be applied to
implement control objectives other than state transfer, including
non-local unitary operations.

We acknowledges funding from \mbox{EPSRC} ARF Grant EP/D07195X/1,
Hitachi and the \mbox{EPSRC QIP IRC}, EU Knowledge Transfer Programme 
MTDK-CT-2004-509223 and NSF Grant PHY05-51164.

%%References go here
\bibliographystyle{prsty}
\bibliography{/home/sonia/archive/bibliography/References,/home/sonia/archive/bibliography/SpinChain}
\end{document}